\documentclass[twocolumn]{aastex631}

\usepackage{amsmath}
\shorttitle{Resolving MBHB evolution with particle splitting}
\shortauthors{Franchini et al.}
\graphicspath{{./}{figures/}}

\begin{document}

\title{Resolving massive black hole binaries evolution via adaptive particle-splitting}

\correspondingauthor{Alessia Franchini}
\email{alessia.franchini@unimib.it}

\author[0000-0002-8400-0969]{Alessia Franchini}
\affiliation{Dipartimento di Fisica ``G. Occhialini", Universit\'a degli Studi di Milano-Bicocca, Piazza della Scienza 3, 20126 Milan, Italy}
\affiliation{INFN, Sezione di Milano-Bicocca, Piazza della Scienza 3, I-20126 Milano, Italy}

\author[0000-0001-6106-7821]{Alessandro Lupi}
\affiliation{Dipartimento di Fisica ``G. Occhialini", Universit\'a degli Studi di Milano-Bicocca, Piazza della Scienza 3, 20126 Milan, Italy}

\author[0000-0003-4961-1606]{Alberto Sesana}
\affiliation{Dipartimento di Fisica ``G. Occhialini", Universit\'a degli Studi di Milano-Bicocca, Piazza della Scienza 3, 20126 Milan, Italy}



\begin{abstract}

The study of the interaction of a massive black hole binary with its gaseous environment is crucial in order to be able to predict merger rates and possible electromagnetic counterparts of gravitational wave signals.
The evolution of the binary semi-major axis resulting from this interaction has been recently debated, and a clear consensus is still missing, also because of several numerical limitations, i.e. fixed orbit binaries or lack of resolution inside the cavity carved by the binary in its circumbinary disc.
Using on-the-fly particle-splitting in the 3D meshless code \textsc{gizmo}, we achieve hyper-Lagrangian resolution, which allows us to properly resolve the dynamics inside the cavity, and in particular for the first time the discs that form around the two components of a live binary surrounded by a locally isothermal gaseous circumbinary disc. 
We show that the binary orbit decays with time for very cold and very warm discs and that the result of the interaction in the intermediate regime is strongly influenced by the disc viscosity as this essentially regulates the fraction of mass contained in the discs around the binary components as well as the fraction that is accreted by the binary. We find the balance between these two quantities to determine whether the binary semi-major axis decreases with time.

\end{abstract}


\section{Introduction} \label{sec:intro}

Massive black holes (MBHs) are expected to reside in the nuclei of (virtually all) massive galaxies \citep[e.g.][]{Kormendy2013}.
When two massive galaxies merge the MBHs in their centers migrate towards the center of the newly formed galaxy owing to the effect of dynamical friction \citep{Chandrasekhar1943} against the background of stars and gas.
The two MBHs are expected to form a bound massive black hole binary (MBHB) at parsec scales. At these scales dynamical friction becomes inefficient and further evolution of the binary orbit towards coalescence requires a mechanism to extract angular momentum and energy from the binary.
The main mechanisms proposed in the literature are three-body scattering of stars intersecting the binary orbit \citep[e.g.,][]{Quinlan1996,sesana2007} or the interaction with a circumbinary gaseous disc \citep{escala2005,dotti2007,Cuadra2009} depending on the binary separation scale (see \citealt{Bortolas2021} for details on the competition between stellar and gaseous interaction). 

Each of the two massive galaxies brings with it a significant amount of gas that sinks to the center forming a circumbinary accretion disc. This gaseous disc might facilitate the MBHB merger and give rise to electromagnetic counterparts of the gravitational wave (GW) emission \citep{begelman1980,ArmitageNatarajan2002,Lodato2009}. 
A coplanar accretion disc can extend down to a few times the binary separation \citep{artymowicz1994}. The disc orbits resonate with the binary orbit at discrete locations (outer Lindblad resonances), leading to the exchange of angular momentum between the disc and the binary \citep{lynden-bell1972,Lin1986}. The magnitude of the resonant torques depends on the binary potential, i.e. its mass ratio and eccentricity, and is proportional to the disc surface density at the resonance locations \citep{goldreich1979}. Therefore, the amount of angular momentum transferred from the binary to the disc at the resonances depends on the disc properties as well. 

Very early numerical simulations \citep{artymowicz1994,artymowicz1996} investigated the interaction of a binary with its gaseous circumbinary disc finding that only a small amount of material is able to enter the cavity carved by the binary and to accrete onto the binary components. The main finding of these works is that the binary semi-major axis decreases with time owing to the interaction with the disc.
This picture has been recently challenged by a few works \citep{Munoz2019,Duffell2019,Munoz2020} employing 2D (and one 3D simulation, see \citealt{Moody2019}) static or moving-mesh grid numerical simulations with fixed binary orbits. In particular, these studies found that the secular angular momentum transfer onto the binary is strongly positive within the range of binary and disc parameters explored. More recently \cite{tiede2020} found, using the same numerical techniques, that the sign of the torque exerted by the disc onto the binary depends on the disc temperature, i.e. on its aspect ratio $H/R$, for locally isothermal discs. In particular, they found discs with $H/R \lesssim 0.04$ to shrink the binary. 
Using 3D smoothed particle hydrodynamics (SPH) simulations of locally isothermal discs, \cite{heathnixon2020} found instead the threshold value for binary expansion to be $H/R\simeq 0.2$.
Other works that employed SPH simulations in the regime where the disc self-gravity cannot be neglected, and the disc temperature changes with time, found binary shrinking as a result of the interaction with massive discs regardless of the disc temperature \citep{Cuadra2009,roedig2012,franchini2021}.
The discrepancy in the results inferred using different numerical techniques has been argued to originate from the lack of numerical resolution in SPH simulations which, in constrast with grid-based ones, were unable to properly resolve the dynamics of the gas streams entering the cavity, artificially suppressing the positive torque associated to such gas that forces the binary to expand \citep{Munoz2019}.

In this letter, we employ the public version of the code {\sc gizmo} \citep{Hopkins2015} in its mesh-less finite mass (MFM) mode, coupled with adaptive particle-splitting for numerical refinement of the gas dynamics inside the disc cavity, in the aim at accurately measuring the gravitational and accretion torques that originate from the discs that form around the binary components (also called {\it mini-discs}) onto the binary itself.\footnote{ 
The adaptive particle splitting technique has been previously successfully employed in isolated \citep{Curtis2015} and cosmological \citep{AA2020} simulations to resolve the dynamics of the gas as it approaches the central massive black hole.}

In Section \ref{sec:sim} we describe the details of the numerical method and present the results of our simulations in Section \ref{sec:results}. We discuss the relevance of the parameters explored in Section \ref{sec:params} and finally draw our conclusions in Section \ref{sec:concl}.

\section{Numerical setup} 
\label{sec:sim}

The initial conditions for this work consist of a live binary surrounded by a circumbinary gaseous disc. The 3D distribution of the $10^6$ equal mass gas particles sampling the disc and the initial orbit of the two sink particles of the binary are generated using the SPH code {\sc phantom} \citep{phantom2017}. 
The equal mass binary has an initial mass $M=M_1+M_2=1$ and a separation $a=1$. The circumbinary disc initially extends from $R_{\rm in} =2a$ to $R_{\rm out}=10a$, with a fixed aspect ratio $H/R=0.1$. The gas is described by a locally isothermal  equation of state with the sound speed $c_{\rm s}$ defined by Eq.~(4) in \cite{farris2014}, and a surface density profile scaling as $\Sigma \propto R^{-3/2}$, normalised to get a total mass $M_{\rm disc}=0.1M$. 

The simulations are performed with the MFM method in {\sc gizmo}, keeping the binary live, i.e. its orbit is allowed to change during the evolution.
The MFM is a mesh-free Lagrangian approach which encapsulates the advantages of both grid-based and particle-based codes. In particular, by partitioning the domain with discrete tracers (particles/cells), and solving the integral form of the fluid dynamics equations via a finite-volume Godunov method, this numerical technique allows to obtain intrinsic adaptivity in resolution while minimizing the numerical advection of angular momentum through the `cells' and also ensures shock capturing without the need of the extra artificial viscosity terms commonly used in SPH codes.
We here also include the effect of gas viscosity entering the Navier-Stokes fluid equations as described in \citet{hopkins2016}, assuming a shear viscosity in the disc $\nu=\alpha c_{\rm s}H$ parametrised using a viscosity parameter $\alpha=0.1$ \citep{ss1973}, and no bulk viscosity. 

\subsection{Gas accretion and gravitational forces}

Every time a gas particle approaches one of the sinks, entering its sink radius $r_{\rm sink}$, we flag it as eligible to be accreted. During the accretion event, conservation of mass, and linear and angular momentum are ensured, in the same way it is done in the {\sc phantom} code \citep{bate1995}.\footnote{Note that in the public implementation of the accretion in \textsc{gizmo}, the linear momentum conservation only accounts for the change in the centre-of-mass position and velocity, neglecting the acceleration term \citep[see][for details]{price2017}. Since this has the effect of reducing the momentum conservation accuracy, which is crucial in our study, we suitably added it to the code.}
In this work, we neglect the disc self-gravity, and only include the gravitational interaction between i) the two sink particles, and ii) sink and gas particles \citep[see][for a discussion about the role of self-gravity]{franchini2021}. In order to get a high accuracy in the dynamical evolution, we suitably modify the code as follows. First, we include the gravitational acceleration exerted by the gas particles onto the two black holes via direct summation, which guarantees the most accurate estimation of the acceleration on both sinks. 
Secondly, we introduce a new timestep criterion for the sink particles that forces both of them to evolve on the shortest common timestep. 
Finally, we re-set the centre of mass and the centre of mass velocity of the entire system to the origin at each course timestep, in order to avoid the build-up of small conservation errors in the linear momentum that, over the long integration times we consider, might displace the binary from the center. Note that this does not influence the dynamics of the system, since it simply corresponds to a rigid shift of the positions and velocity of the centre of mass of all sink and gas particles to the origin.

\begin{figure}
    \centering
    \includegraphics[width=\columnwidth]{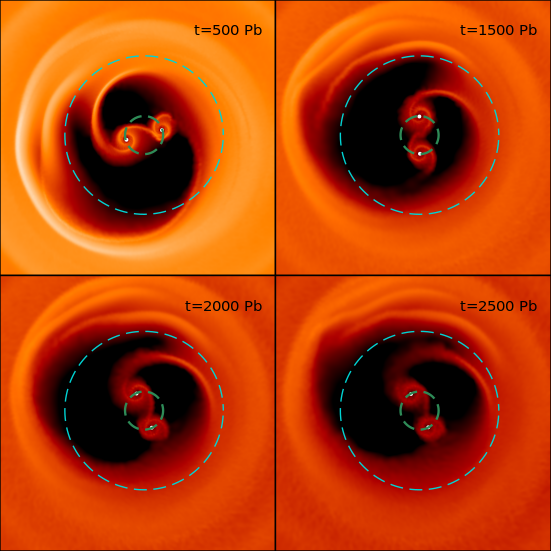}
    \caption{Column density plots of the circumbinary disc and the discs around the binary components (shown by the white circles).  The view is of the $x$-$y$ plane (i.e. the binary orbital plane) and the density has been integrated through $z$. The color scale spans about three orders of magnitude in density and is the same for all the plots. The green dashed circle corresponds to the initial binary orbit, and the cyan dashed circle to the location of the strongest Lindblad resonance, i.e. $2.08a$.}
    \label{fig:maps}
\end{figure}

\subsection{Particle splitting}

In order to increase the resolution when and where necessary, without globally slowing down our simulations, we employ an on-the-fly adaptive particle-splitting approach, which is similar in spirit to the adaptive mesh refinement of grid-based codes. Such technique represents a natural generalisation of the refinement/de-refinement scheme in {\sc arepo} \citep{Springel2010} and {\sc gizmo} \citep{Hopkins2015} to maintain an almost constant mass per cell during the simulations when a finite-volume scheme is employed.
In this work, we exploit the particle splitting algorithm in {\sc gizmo}, splitting gas particles that enter a sphere of radius $r_{\rm ref}$ centred on the centre of mass of the binary-disc system. In particular, particles at radii $r<r_{\rm ref}$ are split when their mass $m\gtrsim \chi_{\rm ref}\,m_{\rm max}$, with $m_{\rm max}$ a scale mass and $\chi_{\rm ref}$ the refinement factor, a suitably defined function, which we choose of the form 
\begin{align}
\chi_{\rm ref}(r) = \left\{ \begin{array}{cc} 
                \chi_{\rm ref}^{\rm min} & \hspace{1mm} \xi<\xi_{\rm min}  \\
                A\xi^2+B\xi + C & \hspace{1mm} \xi_{\rm min} \leq \xi\leq \xi_{\rm ref} \\
                1 & \hspace{1mm} \xi>1  \\
                \end{array} \right.
                \label{eq:reffunc}
\end{align}
Here, $\xi=r/r_{\rm ref}$, $\xi_{\rm min}=r_{\rm min}/r_{\rm ref}$ with $r_{\rm min}$ the radius at which the maximum resolution is reached, $\chi_{\rm ref}^{\rm min}=1/32$ and the coefficients are defined as 
$A=(1-\chi_{\rm ref}^{\rm min})/(1-\xi_{\rm min})^2,\,B=-2A\xi_{\rm min},\,C=1-A(1-2\xi_{\rm min})$.
In particular, we choose $r_{\rm min}=0.75a$, that guarantees that the flow around the two components of the binary is always at the maximum resolution, and $r_{\rm ref}=4a$, that allows to properly resolve the edge of the cavity and the dynamics of the gas streaming inwards. 


When a particle is flagged for refinement, two children are created, each with mass $0.5\,m$, and located at a distance $dr={\rm min}(0.25\,h,0.35\,d_{\rm nbg})$, where $h$ is the particle kernel size and $d_{\rm nbg}$ is the distance to the nearest neighbor, from the position of the parent and on opposite sides along a random direction. The numerical factors are necessary to minimize perturbations to hydrodynamic quantities and avoid overlapping of fluid elements \citep{AA2020}. All the remaining properties are instead directly inherited by the children. This approach allows the almost perfect conservation of momentum, angular momentum and energy throughout the simulation.
The density is re-calculated immediately after the splitting using the standard approach.
In addition to splitting, particles are also allowed to merge when their mass is significantly below the resolution requirements, i.e. $m<\chi_{\rm ref}\,m_{\rm min}$, with $m_{\rm min}$ the scale mass for merger. In our numerical simulations, we set $m_{\rm max}$ equal to the initial mass of the particles in the disc, and $m_{\rm min}$ to 1/1000th of it, which translates in neglecting merging, although we also explored different values and the inclusion of mergers, finding negligible differences.  

\section{Results} \label{sec:results}

We now describe our main results, exploring how different parameters and resolution requirements affect our conclusions.

\subsection{Lagrangian resolution}

As a first check of our numerical and physical setup, we performed a simulation using standard Lagrangian resolution, i.e. switching off particle splitting, and a large sink radius $r_{\rm sink}=0.2$, in order to obtain a fair comparison with the results presented in \cite{heathnixon2020}. 
We find a general agreement between our simulation with $H/R=0.1$ and that represented by the red dashed curve in Fig. 4 of \cite{heathnixon2020}, with the binary shrinking over time. We also performed the same test with a thicker disc, i.e. $H/R=0.2$, in this case finding binary expansion, as already found by \citet{heathnixon2020}.

We note that the rate at which the binary shrinks (or expands) is in general slightly higher in our runs than in \citet{heathnixon2020}, likely because of the different treatment of hydrodynamics (e.g. different numerical treatment of shocks) in the two codes and the slightly different equation of state employed for the gas. However, a precise comparison between the codes is beyond the scope of this work. 

\subsection{Hyper-Lagrangian resolution}
\label{sec:thick}

The use of pure Lagrangian resolution has been argued to possibly lead to a less accurate treatment of the interaction between a binary and its circumbinary disc, because of the small amount of particles entering the cavity, resulting in a poor resolution around the two MBHs. In order to overcome this potential limitation, we employ here on-the-fly particle splitting which allows us to significantly increase the resolution of our simulations in the cavity at a moderate computational cost.

We here show and discuss the result of our fiducial case with particle splitting (implemented according to Eq.~\ref{eq:reffunc}), disc aspect ratio $H/R=0.1$, constant throughout the disc, and viscosity parameter $\alpha=0.1$. 
Compared to the purely Lagrangian tests, the sink size is now reduced to $r_{\rm sink}=0.05$, to ensure that the gas dynamics around each MBH is properly resolved.

Figure \ref{fig:maps} shows the column density plots of the circumbinary disc together with the resolved discs around the two binary components (white circles) obtained with particle splitting at $t=500,1500,2000,2500\,P_{\rm b}$. We can clearly resolve the dynamics inside the cavity carved by the live binary using our hyper-Lagrangian approach. The cavity does not remain circular, as already observed in previous studies \citep[][]{ragusa2016,Munoz2019,heathnixon2020}, and the formation of shocks at the cavity edge is also expected due to the part of the streams that gets flown back to the disc inner edge after entering the cavity.

We test the conservation of angular momentum in the simulations with and without particle splitting, following the procedure outlined in \cite{franchini2021} \citep[see also][]{Munoz2019}, in order to ensure that the particle splitting algorithm does not significantly affect the conservation. We divide the contribution of the disc to the change in the binary angular momentum in a gravitational and an accretion part, the latter including also the fraction of the accreted angular momentum that is converted into the spin of the two MBHs.

\begin{figure*}
    \centering
    \includegraphics[width=\columnwidth]{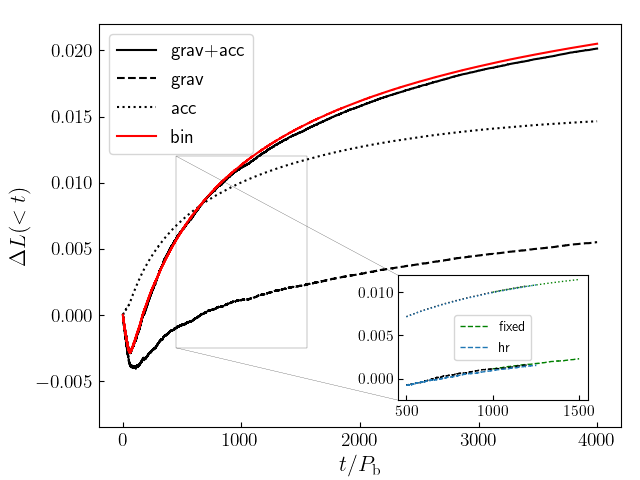}
    \includegraphics[width=\columnwidth]{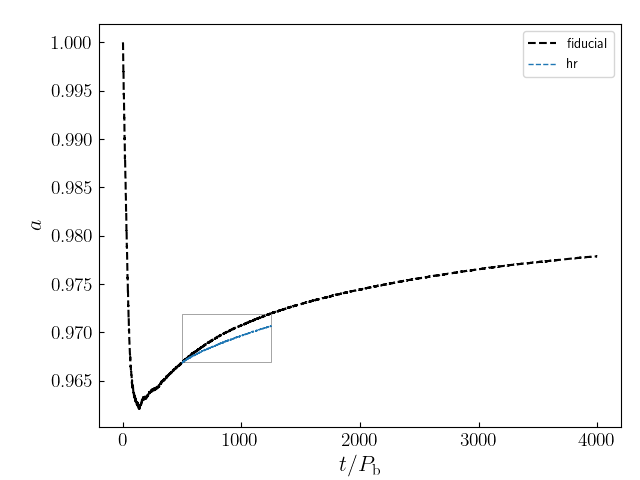}
    \caption{Left panel: conservation of angular momentum in our fiducial run with particle splitting. The dotted and dashed curve show the cumulative contribution of mass accretion and gravity respectively and the black straight line is the sum of the two. The red line shows the evolution of the angular momentum of the live binary. The inset shows the accretion (dotted line) and gravitational (dashed line) torques measured from our fixed binary orbit (green) and high resolution (blue) run.}  Right panel: Evolution of the binary semi-major axis with time in our fiducial (black line) and higher resolution (blue line) run with particle splitting. 
    The time is in units of the binary orbit period, i.e. $P_{\rm b}=2\pi$ in code units.
    \label{fig:angmom}
\end{figure*}

The left panel of Figure \ref{fig:angmom} shows the contributions to the binary angular momentum change as computed directly from our fiducial simulation. In particular, from the comparison between the black straight line (i.e. sum of gravitational and accretion contribution) and the red line that shows the live binary angular momentum change, it is evident that the angular momentum is conserved in our fiducial run almost exactly. The small discrepancy is due to our approximate calculation of the gravitational torque between two subsequent snapshots instead of calculating it at each timestep. 
As a further consistency check, we also compared the same quantities in a simulation without particle splitting (keeping $r_{\rm sink}=0.05$), finding the impact of the refinement scheme on the conservation to be negligible.

Since the simulation conserves angular momentum, we can write the evolution of the binary semi-major axis as

\begin{equation}
    \frac{\dot{a}}{2a} = \frac{\dot{L}_{\rm z,acc}}{L_{\rm z}} + \frac{\dot{L}_{\rm z,grav}}{L_{\rm z}} - \frac{\dot{\mu}}{\mu} - \frac{\dot{M}}{2M} + \frac{e\dot{e}}{(1-e^2)}
    \label{eq:adot}
\end{equation}
where $L_{\rm z}=\mu\sqrt{GMa(1-e^2)}=L_{\rm z,grav}+L_{\rm z,acc}$ is the $z$ component of the binary angular momentum, $\mu$ is the reduced mass and $e$ is the binary eccentricity. 
The first term is always positive and represents the accretion of angular momentum onto the binary (dotted line in Fig. \ref{fig:angmom}). The second term is given by the sum of the positive contribution of the discs around the two components and the negative contribution of the circumbinary disc and corresponds to the dashed line in Fig. \ref{fig:angmom}. The terms due to the accretion of mass (i.e. third and fourth terms) are additional negative contributions to the semi-major axis evolution. The last term comes from the binary eccentricity evolution and we find it to be negligible in our runs.
Therefore we essentially have two positive terms whose effect is to drive the binary apart and two negative contribution that remove angular momentum from the binary driving it towards merger.
In more physical terms, the angular momentum change due to the gravitational and accretion torques translates into a change of the different elements of the system, i.e. the masses and orbital elements. Since the eccentricity contribution is negligible in our runs, the evolution of the binary separation depends on the fraction of the angular momentum exchanged with the disc goes into the mass accretion terms.

The right panel of Figure \ref{fig:angmom} shows the evolution of the live binary semi-major axis with time, directly calculated from our fiducial run using energy and angular momentum conservation. 
The binary semi-major axis decreases within the first $\sim$100 binary orbits, due to a transient phase as the disc adjusts to an equilibrium configuration. 
After the first 100 orbits, the shape (and size) of the cavity changes significantly as the streams are flown back to the circumbinary disc. This causes the amount of material at the 2:1 Lindblad resonance (represented by the cyan dashed circle in Fig. \ref{fig:maps}) to decrease by a factor of $\sim$5-10 weakening the strength of the resonance. As a consequence, the positive torque provided by the discs around the binary components, which is able to overcome the negative contribution of the circumbinary disc, leads to the increase of the binary separation. \footnote{ As a further check of our numerical setup, we ran a very high Lagrangian resolution (i.e. employing $10^7$ SPH particles) simulation using the {\sc phantom} code, finding the same behaviour of our fiducial run. }
We find the expansion phase to be very mild and the semi-major axis to increase at a progressively slower pace. This might be due to the progressive decrease of the disc mass owing to accretion, which drops to $M_d\approx 0.04M$ after $4000$ orbits.
We note that this phase appears to be in broad agreement with the findings of grid code simulations \citep{Munoz2019,Duffell2019} indicating that indeed resolving the dynamics inside the cavity is important in order to properly investigate the binary-disc interaction. 

\subsection{The impact of spatial/mass resolution and of live/fixed binary orbits}

 Although the extremely high resolution achieved owing to the hyper-Lagrangian refinement enabled us to better resolve the gas flowing into the cavity, it might be argued that the resolution achieved is insufficient to properly measure torques in the cavity.  We have therefore estimated the average inter-particle separation of the gas in the discs surrounding the binary components in our fiducial simulation, finding $\Delta r \simeq 0.015a$. This corresponds to, e.g., the mid-resolution run in \cite{tiede2020}, already confirming the good accuracy of our measured torques. Nonetheless, we have further investigated the impact of resolution inside the cavity by decreasing the minimum refinement factor down to $\chi_{\rm ref}^{\rm min}=1/64$, i.e. by a factor 2. In order to avoid the initial transient in the orbital evolution, we have restarted our fiducial run from the 500th orbit, carefully emptying the cavity to prevent numerical instabilities arising from a sudden increase in resolution near the binary components. Note that the amount of mass removed by this process is negligible compared to the disc mass, and the removal does not affect the subsequent evolution, since the discs around the two components reform in less than one orbit. We find the structure and density of these discs to remain essentially unaltered and unaffected by the increase in resolution. The blue lines in the left panel inset of Figure \ref{fig:angmom} show the gravitational and accretion torque measured in this case.  We find the accretion torque to remain the same while the gravitational torque is slightly lower in this high resolution run. This reflects on the evolution of the binary semi-major axis (see the blue line in the right panel inset in Fig. \ref{fig:angmom}) which increases at a slightly slower pace.

The majority of previous studies on circumbinary discs is based on simulations in which the binary is kept on a fixed orbit. While this assumption might be reasonable when the change in binary properties is small, a proper analysis of its impact is in order. For this reason, we have also performed a simulation identical to our fiducial run, in which we kept the binary orbit fixed, as in previous studies \citep{Munoz2019,Duffell2019,Moody2019}. 
The comparison between the fixed and live binary orbit run, in terms of gravitational and accretion torques, is shown by the green lines in the left panel inset of Figure \ref{fig:angmom}. We started the simulation from the 1000th binary orbit of our fiducial live binary run since, by this time, the disc is in a quasi-steady state. We find that the accretion and gravitational torques do not vary significantly compared to our fiducial live binary simulation. We note however that the surface density of the discs that form around the two binary components is slightly lower in the simulation with fixed binary orbit. This does not cause the torques to be significantly different over a few hundred orbits but it is an effect that will be further analysed in a future work.  


\subsection{Disc temperature}
\label{sec:thicker}

In order to determine whether there is a critical condition that discriminates between binary shrinking and expansion, we also explored different disc temperatures, keeping the small sink radius $r_{\rm sink}=0.05$ and using particle splitting.

First, we decreased the disc aspect ratio to $H/R=0.03$, i.e. within the regime where both grid-code \citep{tiede2020} and SPH \citep{heathnixon2020} simulations find binary shrinking, and we found our simulations with particle splitting to be in agreement with previous studies. We note that, given the low disc temperature, the mass in the discs around the binary components is very low regardless of the refinement and this reduces the positive contribution to the gravitational torque.

\cite{heathnixon2020} found, using pure SPH simulations without particle splitting, that circumbinary discs with aspect ratios $H/R\geq0.2$ cause the binary orbit to expand owing to the large amount of material that is able to enter the cavity and to accrete onto the binary.
For completeness, and a further check for our fiducial run, we also explored the case of a thicker disc with $H/R$=0.2.
\footnote{Note that \cite{heathnixon2020} used a slightly different equation of state for the disc. In particular their disc is flared with $H/R\,(R_{\rm in})=0.2$ while in our runs the aspect ratio is constant throughout the disc.}

\begin{figure*}
    \centering
    \includegraphics[width=0.75\columnwidth]{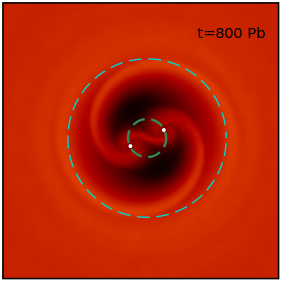}   
    \includegraphics[width=\columnwidth]{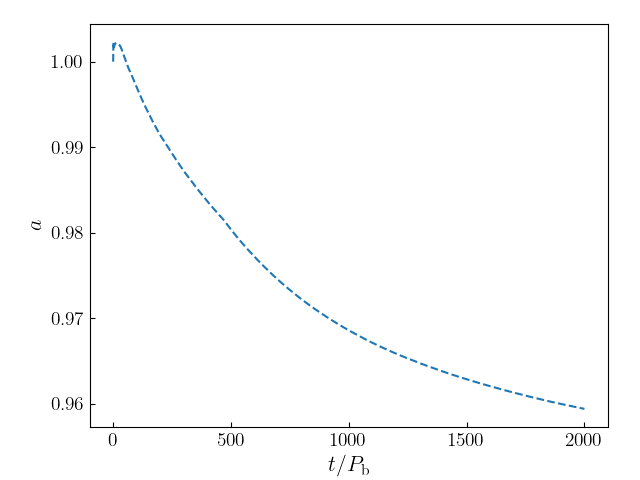}
    \caption{Left panel: Same as Fig. \ref{fig:maps} but with $H/R=0.2$ and shown at $t=800 P_{\rm b}$. The density scale is the same as in Fig. \ref{fig:maps}. Right panel: Same as lower panel of Fig. \ref{fig:angmom} but with a disc aspect ratio $H/R=0.2$.}
    \label{fig:maps2}
\end{figure*}

The left panel of Figure \ref{fig:maps2} shows the column density plots of the circumbinary disc together with the resolved discs around the two binary components (white circles) obtained with particle splitting for a thick accretion disc. The density scale is the same as in Fig. \ref{fig:maps}. We find that the material that enters the cavity is significantly larger compared to the thinner disc simulation.

We can see from the right panel of Figure \ref{fig:maps2} that the binary semi-major axis decreases with time, after undergoing a very brief and mild expansion within the first 20 orbits, even though the disc thickness is in the regime \cite{heathnixon2020} found for expansion. 
Despite the gravitational and accretion torques being positive throughout the duration of the simulation, we find the negative contribution of the accretion of mass (and not angular momentum)  to dominate the evolution of the binary.

\section{Discussion}
\label{sec:params}
Interestingly we find the binary to shrink for thin and very thick discs and that there is a regime of intermediate disc thickness where the negative and positive torques balance out, leading to a phase of possible binary expansion. 

We compared our simulation with $H/R=0.1$ with and without particle splitting,  and found that having a small sink radius without resolving the discs that form around the binary components might result in an overestimate of the (positive) gravitational torque and this in turn leads to a much faster binary expansion phase. This lasts until the accretion of mass onto the binary components becomes large enough to overcome the positive torque and is very sensitive to the resolution inside the cavity. 
In our fiducial case the expansion is significantly milder and slower, however it seems to last for much longer due to the positive torque from the discs around the two components being strong enough to balance the negative torque from the accretion of mass.
In our low resolution simulation we find the binary semi-major axis to decrease again after $\sim 800$ orbits.
We further explored the effect of resolution by increasing the resolution inside the cavity by a factor 2 with respect to our fiducial run. We find that our results are overall not affected by the numerical resolution. 

The balance between the positive and negative torques is essentially regulated by the disc viscosity, which determines the amount of material that enters the cavity, forms the discs and then accretes onto the binary components. Therefore different treatments and different $\alpha$ values are very likely to change the binary evolution. 
For instance, decreasing the value of $\alpha$ by a factor two in our fiducial run with aspect ratio $H/R=0.1$ leads to the binary semi-major axis decreasing with time without transiting to the expansion phase.

Since the MFM method has been proven to give more accurate results on standard tests compared to the pure SPH approach \citep{Hopkins2015}, even with a small number of neighbours, we reran our fiducial simulation with particle splitting, but employing only $32$ neighbours instead of the $58$ used in the rest of this work.
We find the expansion phase to be significantly milder.

We also note that, by $3500\,P_{\rm b}$, the binary in our fiducial run has already accreted $\sim 60\%$ of the initial disc mass and the semi-major axis is still increasing, even though at a progressively slower pace.
In order to understand whether this phase is just a transient, in a future work we will implement gas particle injection to provide a constant source of gas into the circumbinary disc to feed the discs inside the cavity. 

Finally, we do not find binary expansion in the simulation presented in Section \ref{sec:thicker} with $H/R=0.2$ using particle splitting with a small accretion radius. The reason for this change resides in the gas distribution in the cavity. 
The gas distribution inside the discs around the binary components in the thicker disc case is uniform and therefore removing the inner regions by enlarging the accretion radius leads to a higher specific angular momentum being transferred to the binary during accretion, which in turn causes the binary to expand.
Although whether there exist a threshold for binary expansion in terms of the disc aspect ratio still remains to be determined, such large disc thicknesses are well beyond the typical aspect ratio we may expect for circumbinary discs surrounding massive black hole binaries, if their structure resembles that of AGN discs \citep{Collin1990}.

\section{Conclusions} \label{sec:concl}

In this work, we use hyper-Lagrangian refinement to resolve for the first time the gas dynamics inside the cavity carved by a live binary in its circumbinary disc, using 3D hydrodynamical simulations employing the mesh-less finite mass numerical algorithm of the code \textsc{gizmo}.
In particular, resolving the discs that form around the binary components allows us to better measure the torques they exert on the binary. 
We find the gravitational torques by these discs to be positive, in broad agreement with the results of previous grid-based simulations \citep{Munoz2019,Duffell2019}. However, we find these not to be strong enough to overcome the negative circumbinary disc torque and the negative angular momentum variation due to the accretion of mass both in the case of thin and very thick discs.

We therefore conclude that the binary semi-major axis typically decreases with time as a result of the interaction with the gaseous accretion disc, if the disc is thin, but even if the disc is very geometrically thick.
Interestingly, we find an intermediate regime where the positive torque exerted by the discs around the binary components is able to counter balance the negative torques, leading to a possibly temporary phase of expansion. However, we caution that this is very sensitive to the viscosity treatment, and therefore worth investigating more in detail in a follow-up paper.

\begin{acknowledgments}
We thank Daniel Price for providing the {\sc phantom} code for numerical simulations and acknowledge the use of {\sc splash} \citep{Price2007} for the rendering of the figures.
We thank Phil Hopkins for providing the {\sc gizmo} code for numerical simulations. 
AF and AS acknowledge financial support provided under the European Union’s H2020 ERC Consolidator Grant ``Binary Massive Black Hole Astrophysics" (B Massive, Grant Agreement: 818691).
AL acknowledges funding from MIUR under the grant PRIN 2017-MB8AEZ.
\end{acknowledgments}

%








\bibliography{references}{}
\bibliographystyle{aasjournal}



\end{document}